\documentclass[a4paper]{article}
\usepackage{amsmath}

\newcommand{\si}{\ensuremath{\sigma_{1}}}
\newcommand{\sj}{\ensuremath{\sigma_{2}}}
\newcommand{\sk}{\ensuremath{\sigma_{3}}}
\newcommand{\half}{{\textstyle \frac{1}{2}}}
\newcommand{\qrt}{{\textstyle \frac{1}{4}}}
\newcommand{\dt}{\! \cdot \!}
\newcommand{\lra}{\ensuremath{\leftrightarrow}}
\newcommand{\ket}[1]{\ensuremath{| #1 \ra}}
\newcommand{\bra}[1]{\ensuremath{\la #1 |}}
\newcommand{\braket}[2]{\ensuremath{\la #1 | #2 \ra}}
\newcommand{\la}{\langle}
\newcommand{\ra}{\rangle}
\newcommand{\clg}{{\mathcal{G}}}
\newcommand{\clh}{{\mathcal{H}}}
\newcommand{\clo}{{\mathcal{O}}}
\newcommand{\bP}{\mbox{\boldmath $P$}}

\newcommand{\spinor}[2]{\ensuremath{
\begin{pmatrix}
#1 \\
#2 \\
\end{pmatrix}}}

\begin{document}

\begin{center}

{\bf\Large Analysis of 1 and 2 Particle Quantum Systems \\
using Geometric Algebra}

\vspace{0.4cm}

Rachel Parker\footnote{e-mail: \texttt{rfp23@mrao.cam.ac.uk}}  and
Chris Doran\footnote{e-mail: \texttt{c.doran@mrao.cam.ac.uk},  
\texttt{http://www.mrao.cam.ac.uk/$\sim$cjld1/}} 

\vspace{0.4cm}

Astrophysics Group, Cavendish Laboratory, Madingley Road, \\
Cambridge CB3 0HE, UK.

\vspace{0.4cm}

\begin{abstract}
When two or more subsystems of a quantum system interact with each
other they can become entangled.  In this case the individual
subsystems can no longer be described as pure quantum states. For
systems with only 2 subsystems this entanglement can be described
using the Schmidt decomposition.  This selects a preferred orthonormal
basis for expressing the wavefunction and gives a measure of the
degree of entanglement present in the system.  The extension of this
to the more general case of $n$ subsystems is not yet known. We
present a review of this process using the standard representation and
apply this method to the geometric algebra representation.  This
latter form has the advantage of suggesting a generalisation to $n$
subsystems.
\end{abstract}

\end{center}

\section{Introduction}
\label{sect:intro}

Quantum entanglement in 2-particle systems is currently well
understood (for a useful review, including an extensive list of
references, see~\cite{lomo01}).  But the quantum behaviour of
many-particle systems is more complicated and less well understood,
and it is these systems that are of interest experimentally. The main
limitation to the theoretical understanding of such systems is that
the techniques which have been developed to analyse 2-particle systems
do not easily generalise. Geometric algebra has the advantage
that the number of particles being analysed dictates the size of the
space but otherwise does not alter the analysis used. In this way,
results developed in simple cases (such as the 2-particle system) can
be more easily generalised to the $n$-particle case.

In this paper we focus on 2-state quantum systems in the cases of one
and two particles.  We start by reviewing the standard matrix-based
approach to single and two-particle pure states.  We describe the
\textit{Schmidt Decomposition}, which provides a measure of the degree
of entanglement present in a given system.  We then introduce the
density matrix to describe both pure and mixed states in a unified
manner.  We next turn to an analysis of the same systems using the
multiparticle spacetime algebra framework developed by Doran, Lasenby
and Gull~\cite{DGL93-states,DGL95-elphys,SLD99}.  As a simple
application we review the properties of the spin singlet state,
frequently encountered in discussions of the Bell inequalities and
EPR-type experiments~\cite{bell-spk}.

\section{Single-Particle Pure States}
\label{sect:1ps}

If there is only one particle present in the system then the spinor,
$\ket{\psi} \in \clh$, can always be written in the form
\begin{equation}
\ket{\psi} = c_0\ket{0} + c_1\ket{1}
\label{psispinor}
\end{equation}
where \ket{0} and \ket{1} are some pair of orthogonal basis states and
$c_0$ and $c_1$ are complex coefficients.  Alternatively, all
information about the state of the particle can be expressed in a
polarisation (or spin) vector, $\bP$, whose components are given by
\begin{equation} 
P_i = \la \hat{\sigma}_i\ra =
\bra{\psi}\hat{\sigma}_i\ket{\psi}/\braket{\psi}{\psi} 
\end{equation}
where the $\hat{\sigma}_i$ are the Pauli matrices
\begin{equation}
\hat{\sigma}_1 = \begin{pmatrix} 0 & 1 \\ 1 & 0 \end{pmatrix}, \quad
\hat{\sigma}_2 = \begin{pmatrix} 0 & -i \\ i & 0  \end{pmatrix}, \quad
\hat{\sigma}_3 = \begin{pmatrix} 1 & 0 \\ 0 & -1  \end{pmatrix}.
\label{paulis}
\end{equation}
It follows from this definition that $|\bP | = 1$.  In this way the
spin state of the particle can be expressed graphically as a point on
the 2-sphere, and any evolution of the state of the particle can be
thought of as a rotation of the polarisation vector.  In many
applications this sphere is known as the \textit{Bloch sphere}.

\section{2-Particle Systems}
\label{sect:2ps}

Suppose that two particles are described by states belonging to
individual Hilbert spaces $\clh_1$ and $\clh_2$.  The joint Hilbert
space for the interacting system is $\clh_1\otimes\clh_2$, consisting
of complex superpositions of tensor products of states in the
individual spaces.  A basis for $\clh_1\otimes\clh_2$ is constructed
by taking the tensor products of the basis vectors for $\clh_1$ and
$\clh_2$.  Therefore any pure state of the composite system,
$\ket{\psi} \in \clh_1\otimes\clh_2$, can be expressed as
\begin{equation} 
\ket{\psi} = \sum_{i,j=0,1} c_{i,j} \ket{i}\otimes \ket{j} 
\equiv  \sum_{i,j=0,1} c_{i,j} \ket{i,j}.
\label{psi}
\end{equation} 
If more than one of the $c_{i,j}$'s are non-zero then each subsystem
is no longer in a pure state and the system is entangled. In order to
quantify the degree of entanglement we re-express equation~(\ref{psi}) in
the form
\begin{equation} 
\ket{\psi} = \cos(\alpha/2) \ket{0',0'} + \sin(\alpha/2) \ket{1',1'},
\end{equation} 
which can always be done via a suitable change of basis. Here,
\ket{0'} and \ket{1'} are orthonormal vectors in $\clh_1$ and $\clh_2$
and $0 \le \alpha \le \pi/2$ (so that $\cos(\alpha/2) >
\sin(\alpha/2)$). Then \ket{0',0'} can be thought of as the separable
state `closest' to \ket{\psi} and $\alpha$ as the degree of
entanglement present in the system.  The procedure by which this basis
is constructed is the \textit{Schmidt Decomposition}, which we now
describe.

\subsection{Schmidt Decomposition}
\label{sd}

An arbitrary wavefunction \ket{\psi} can be rewritten as a sum of two
state vectors through a transformation of the basis vectors in the
following way~\cite{ekert95}.  Let \ket{u} and \ket{v} be unit vectors
of the first and second particles respectively. Define $M$ by
\begin{equation} 
\label{meq}
M=\braket{u,v}{\psi}.
\end{equation} 
$|M|^2$ is nonnegative and bounded so it attains its maximum,
$|M_1|^2$, for some \ket{u_1} and \ket{v_1}. The choice of \ket{u_1}
and \ket{v_1} is not unique since they can at most be determined only
up to phase and it is possible that there are other degeneracies as
well.

Let \ket{u'} be any state of the first particle which is orthogonal to
\ket{u_1}, and let $\epsilon$ be an arbitrarily small number. Then 
\begin{equation} 
\left|\,\ket{u_1}+\epsilon\ket{u'}\right|^2 = 1 + \clo(\epsilon^2)
\end{equation}
so that up to order $\epsilon^2$, $\ket{u_1}+\epsilon\ket{u'}$ is a
unit vector.  We then find that
\begin{equation} 
\braket{u_1+\epsilon u',v_1}{\psi} = M_1 +\epsilon\braket{u',v_1}{\psi}
\end{equation} 
so
\begin{equation} 
\left| \braket{u_1+\epsilon u',v_1}{\psi} \right|^2 = |M_1|^2 +
2\mbox{Re}(\overline{M_1}\epsilon\braket{u',v_1}{\psi}) +
\clo(\epsilon^2). 
\end{equation}
But \ket{u_1} was chosen so that the scalar product in Eq.~(\ref{meq})
is a maximum. Therefore we must have that
\begin{align}
|M_1|^2 + 2\mbox{Re}(\overline{M_1}\epsilon\braket{u',v_1}{\psi}) &
\le  |M_1|^2, \nonumber \\ 
\implies \quad \mbox{Re}(\overline{M_1}\epsilon\braket{u',v_1}{\psi})
& \le  0. 
\label{ineq}
\end{align}
The choice of phase of \ket{v_1} is arbitrary, however, so to ensure
that Eq.~(\ref{ineq}) is satisfied we must have that 
\begin{equation}
\label{uprm}
\braket{u',v_1}{\psi} = 0, \qquad \forall \quad\! u' \in
\clh^{u_1^\bot}_1 = \{u'\in \clh_1 | \braket{u_1}{u'}=0\}. 
\end{equation}
Similarly, we can show that the same restriction applies to the second
particle so that
\begin{equation}
\label{vprm}
\braket{u_1,v'}{\psi} = 0, \qquad \forall \quad\! v' \in
\clh^{v_1^\bot}_2 = \{v'\in \clh_2 | \braket{v_1}{v'}=0\}. 
\end{equation}
If we now define a new wavefunction
\begin{equation} 
\ket{\psi'} = \ket{\psi} - M_1\ket{u_1,v_1}
\end{equation}
then $\ket{\psi'}$ also satisfies Eqs.~(\ref{uprm}) and~(\ref{vprm})
and has the additional property that
\begin{equation} 
\braket{u_1,v_1}{\psi'}=0.
\end{equation}
From this it follows that $\psi' \in \clh^{u_1^\bot}_1 \otimes
\clh^{v_1^\bot}_2$ and we can repeat the above process on
$\psi'$. Importantly, the dimension of $\clh^{u_1^\bot}_1 \otimes
\clh^{v_1^\bot}_2$ is smaller than the dimension of
$\clh_1\otimes\clh_2$ so this process must terminate. We finally
obtain
\begin{equation} 
\ket{\psi} = \sum_i M_i \ket{u_i,v_i},
\end{equation}
where the sum is over the smaller dimension of $\clh_1$ and $\clh_2$
and \{\ket{u_i}\} and \{\ket{v_i}\} are orthonormal sets.

For the case where each subsystem has dimensionality 2 we find that
\begin{equation} 
\ket{\psi} = M_1 \ket{u_1,v_1} + M_2 \ket{u_2,v_2}.
\end{equation} 
The phases of $M_1$ and $M_2$ can be absorbed into \ket{u_1,v_1} and
\ket{u_2,v_2} so we can set them to be real. This decomposition can be
written explicitly as
\begin{align}
\ket{\psi} =& \rho^{1/2} e^{i\chi}\left(
\cos(\alpha/2) e^{i\tau/2}
\begin{pmatrix}
\cos(\theta_1/2)  e^{-i\phi_1/2} \\ \sin(\theta_1/2) e^{i\phi_1/2}
\end{pmatrix}
\otimes 
\begin{pmatrix}
\cos(\theta_2/2) e^{-i\phi_2/2} \\ \sin(\theta_2/2) e^{i\phi_2/2}
\end{pmatrix}
\right. \nonumber \\ 
& \left. + \sin(\alpha/2) e^{-i\tau/2}
\begin{pmatrix}
\sin(\theta_1/2) e^{-i\phi_1/2} \\ -\cos(\theta_1/2) e^{i\phi_1/2}
\end{pmatrix}
\otimes 
\begin{pmatrix}
\sin(\theta_2/2) e^{-i\phi_2/2} \\ -\cos(\theta_2/2) e^{i\phi_2/2}
\end{pmatrix}
\right). 
\label{Schmidt}
\end{align}
This is the Schmidt decomposition for a bipartite 2-state system.  In
writing this we have satisfied the condition that $\cos\!\alpha \ge
\sin\!\alpha$ since otherwise $\left|\braket{u_2,v_2}{\psi}\right|^2 >
\left|\braket{u_1,v_1}{\psi}\right|^2$ which contradicts our choice of
\ket{u_1,v_1}.

\subsection{The Density Matrix}
\label{dm}

If we want to calculate expectation values for one particle only and
the state of the other particle is unknown then clearly we cannot
write down the full wavefunction.  We are forced instead to turn to
the density operator, $\hat{\rho}$, defined by 
\begin{equation} 
\hat{\rho} = \ket{\psi}\bra{\psi} = \sum_{i,j,k,l}
c_{i,j}c_{k,l}^*\ket{i}\bra{k}\otimes\ket{j} \bra{l}   
\label{defrho}
\end{equation} 
In terms of the density operator the expectation value of any
observable $\hat{Q}$ is given by 
\begin{equation}
 \la \hat{Q} \ra = \mbox{tr}(\hat{\rho} \hat{Q}).  
\end{equation}
The density operator for each particle is given by
\begin{equation} 
\hat{\rho}_1 = \mbox{tr}_2\hat{\rho} = \sum_j
\bra{j}\hat{\rho}\ket{j} \qquad
\hat{\rho}_2 = \mbox{tr}_1\hat{\rho} = \sum_i
\bra{i}\hat{\rho}\ket{i}  
\end{equation} 
so that the expectation value for the $i$th particle can be calculated
by 
\begin{equation}
\la Q \ra_i = \mbox{tr}(\hat{\rho}_i Q). 
\end{equation}
For systems entangled with an (unknown) environment the density matrix
represents our ultimate state of knowledge of the system.  This has
important consequences for the interpretation of quantum mechanics.
For a recent review of these ideas, see Paz \& Zurek~\cite{paz00}.

\section{Geometric Algebra}
\label{sect:gar}

Geometric algebra (GA) is essentially Clifford algebra with added
geometric content.  Since Clifford algebras are a fundamental part of
the treatment of 2-state quantum systems (through the description of
quantum spin), we expect that formulating the theory in a GA framework
should bring added geometric insight.  This idea was first explored by
Hestenes in a series of papers dating back to the
sixties~\cite{hes67,hes71,hes751}.  We start by reviewing the
treatment of single-particle systems.  These are described within the
GA of 3D space, denoted $\clg_3$.  As an orthonormal basis for this we
take
\begin{equation}
1, \quad \{ \sigma_k \}, \quad \{ I \sigma_k \}, \quad I = \si\sj\sk.
\end{equation}
The reverse operation (which flips signs of bivectors and trivectors)
is denoted with a tilde, and angle brackets $\la M \ra_k$ are used to
project onto the grade-$k$ part of $M$.  For the projection onto the
scalar part we simply write $\la M \ra$.  For an introduction into the
geometric algebra of 3D space see~\cite{hes-nf1,DL-course}.

\subsection{Single-Particle Systems}

The simplest example of a 2-state system is provided by quantum spin.
Spin states can be represented as complex 2-component vectors known as
\textit{spinors}.  These can be given a more natural encoding within
$\clg_3$ by defining a linear one-to-one map between the state (as a
complex vector) and a multivector constructed from even-grade terms
(scalars and bivectors).  The simplest such mapping is defined
by~\cite{DGL93-states,DGL95-elphys}
\begin{equation} 
\ket{\psi}  = \spinor{a_0+ia_3}{-a_2+ia_1}
\lra \psi = a_0+a_kI\sigma_k ,
\end{equation}
so that the basis elements \ket{0} and \ket{1} map as
\begin{equation}  
\ket{0} \lra 1 \qquad \mbox{and} \qquad \ket{1} \lra
-I\sigma_2. 
\end{equation}
In this way $\psi$ sits inside the space spanned by $\{1,I\sigma_k\},
(k=1,2,3)$. It follows that
\begin{equation}
\psi\tilde{\psi} = (a_0+a_kI\sigma_k)(a_0-a_kI\sigma_k) 
= a_0^2+a_1^2+a_2^2+a_3^2 =\tilde{\psi}\psi\equiv \rho
\end{equation}
where $\rho$ is the scalar magnitude of the state vector.  The
multivector $\psi$ can then be written as
\begin{equation}
\label{rot}
\psi = \rho^{1/2} R.
\end{equation}
$R$ is then an even, normalised multivector in 3-dimensions and so is
a \textit{rotor} --- a generator of rotations.

The action of the Pauli matrices of Eq.~(\ref{paulis}) is
given by
\begin{equation}
\label{sigeq}
\hat{\sigma}_k\ket{\psi} \lra  \sigma_k\psi \sigma_3=-I\sigma_k\psi
I\sigma_3. 
\end{equation}
It follows that multiplication by $i$ is represented by
\begin{equation}
\label{ieq}
i\ket{\psi} = \hat{\sigma}_1 \hat{\sigma}_2 \hat{\sigma}_3 \ket{\psi} 
\lra  \psi I\sigma_3.
\end{equation}
To construct observables we define the inner product of two spinors,
$\psi$ and $\phi$ by
\begin{equation}
\label{inner}
\braket{\psi}{\phi} \lra (\psi,\phi)_s = \la \phi \tilde{\psi} \ra
- \la \phi I\sigma_3 \tilde{\psi} \ra i. 
\end{equation}
As we will see shortly, this definition generalises simply to
multiparticle systems.  From Eq.~(\ref{inner}) the probability density
is 
\begin{equation} 
\braket{\psi}{\psi} \lra (\psi,\psi)_s = \la \psi \tilde{\psi} \ra
- \la \psi I\sigma_3 \tilde{\psi} \ra I\sigma_3.
\end{equation}
But $\psi I\sigma_3 \tilde{\psi}$ reverses to give minus itself, so
it contains no scalar part. This leaves
\begin{equation}
\label{prob}
\braket{\psi}{\psi} \lra (\psi,\psi)_s = \la \psi \tilde{\psi}
\ra. 
\end{equation} 
For the 1-particle case $\psi \tilde{\psi}$ is purely a scalar and is
equal to $\rho$. In the more general case of $n$-particles we cannot
assume that $\psi \tilde{\psi}$ is purely scalar and so
Eq.~(\ref{prob}), suitably normalised, provides the most general
definition for the probability density.

The other observable we can construct is the expectation for the spin
in the $k$-direction. This is given by
\begin{align}
\braket{\psi}{\sigma_k|\psi}/\braket{\psi}{\psi} \, \lra \,\, &
\rho^{-1}(\psi,-I\sigma_k\psi I\sigma_3)_s \nonumber \\ 
=& \rho^{-1}\la-I\sigma_k\psi I\sigma_3 \tilde{\psi} \ra - \rho^{-1}\la
-I\sigma_k\psi I\sigma_3 I\sigma_3 \tilde{\psi} \ra I\sigma_3
\nonumber \\ 
=& -\rho^{-1}I\sigma_k\cdot\la \psi I\sigma_3 \tilde{\psi} \ra_2 -
\rho^{-1}\la I\sigma_k\psi\tilde{\psi} \ra I\sigma_3.
\end{align}
Since $\psi\tilde{\psi}$ is a scalar $\la I\sigma_k\psi\tilde{\psi}
\ra=0$. Also, $\psi I\sigma_3 \tilde{\psi}$ reverses to give minus
itself and has even grade, so is a pure bivector (again in the
multiparticle space we cannot make this assumption).  Therefore, using
Eq.~(\ref{rot}) we can define the polarisation bivector by
\begin{equation}
\label{P}
P= \la \rho^{-1}\psi I\sigma_3 \tilde{\psi} \ra_2= \la R I\sigma_3
\tilde{R} \ra_2 
\end{equation}
so that 
\begin{equation}
\label{Pk}
P_k=\braket{\psi}{\sigma_k|\psi}/\braket{\psi}{\psi}=
-I\sigma_k\cdot\la R I\sigma_3 \tilde{R}\ra_2= -I\sigma_k\cdot P. 
\end{equation}
In this way the spin of the particle can be thought of as a rotation
of the $I\sigma_3$ plane, where the rotation is given by the wave
function of the particle. The expectation value for the polarisation
in the $k$-direction is then simply the component of $P$ in the
$k$-direction.  This was Hestenes' original insight.  A challenge is
to extend these ideas to the multiparticle framework.

As an example, consider the wavefunction employed in the Schmidt
decomposition~(\ref{Schmidt}) 
\begin{equation} 
\ket{\psi} =
\spinor{\cos(\theta/2)e^{-i\phi/2}}{\sin(\theta/2)e^{i\phi/2}} .
\end{equation} 
In our single particle space this becomes
\begin{align}
\psi &= \cos(\theta/2)e^{-\phi I\sigma_3/2}
-\sin(\theta/2)I\sigma_2e^{\phi I\sigma_3/2} \nonumber \\ 
&= e^{-\phi I\sigma_3/2}(\cos(\theta/2) -\sin(\theta/2)I\sigma_2)
\nonumber \\ 
&= e^{-\phi I\sigma_3/2}e^{-\theta I\sigma_2/2} 
\end{align}
and the polarisation vector $P$ is simply
\begin{align}
P = \psi I\sigma_3 \tilde{\psi} 
&= e^{-\phi I\sigma_3/2}e^{-\theta I\sigma_2/2} I\sigma_3  e^{\theta
I\sigma_2/2}e^{\phi I\sigma_3/2}  \nonumber \\ 
&= \sin(\theta) \cos(\phi) I\sigma_1 + \sin(\theta) \sin(\phi)
I\sigma_2 + \cos(\theta) I\sigma_3   
\end{align}

\subsection{2-Particle Systems}

States for 2-particle systems are constructed in the multiparticle
spacetime algebra (MSTA), which is built from $n$-particle
relativistic configuration space.  A basis for this space is provided
by the vectors $\{\gamma_\mu^a \}$ where the superscript labels the
individual particle space.  Vectors from different spaces are
orthogonal and so anticommute.  It follows that bivectors from
different spaces commute, and hence the even subalgebra of the MSTA
contains the tensor product of a set of non-relativistic algebras.
This is precisely the algebra needed to construct a multiparticle
wavefunction.  A basis for a 2-particle wavefunction is provided by
sums and products of the 1-particle basis elements
$\{1,I\sigma_j^1,I\sigma_k^2\}$, where
\begin{equation}
\sigma_k^a = \gamma_k^a \gamma_0^a, 
\qquad I^a = \gamma_0^a \gamma_1^a \gamma_2^a \gamma_3^a, 
\qquad \mbox{(no sum)}.
\end{equation}
Again, the superscript denotes the particle label, and we abbreviate
$I^1 \sigma_k^1$ to $I\sigma_k^1$, \textit{etc}.

Currently our basis set gives $4\times 4=16$ real degrees of freedom,
whereas we should have only 8 for a 2-particle state.  The solution to
this problem is to demand a consistent meaning for the unit imaginary.
In each separate space multiplication by the imaginary corresponds to
right multiplication by $I\sigma_3$. Since our new space has two such
bivectors we require that
\begin{equation}  
\psi I\sigma_3^1 = \psi I\sigma_3^2, \quad
\psi = \psi \half(1-I\sk^1 \, I \sk^2).
\end{equation}  
We therefore define the 2-particle
\textit{correlator}~\cite{DGL93-states,DGL95-elphys}  
\begin{equation} 
\label{E}
E = \half (1-I\sigma_3^1 I\sigma_3^2), \quad E^2=E.
\end{equation}
$E$ is a projection operator and reduces the number of degrees of
freedom by a factor of 2.  The complex structure in the 2-particle
algebra is now defined by the non-simple bivector $J$, where
\begin{equation}
\label{J}
EI\sigma_3^1=EI\sigma_3^2 =\half (I\sigma_3^1 + I\sigma_3^2)\equiv J.
\end{equation}
The two particle spinor \ket{\psi,\phi} is now mapped to the
multivector
\begin{equation} 
\ket{\psi,\phi} \lra \psi^1\phi^2 E,
\end{equation}
where the superscripts again denote which space the multivector
inhabits.  The result of the action of the unit imaginary becomes
\begin{equation}
i\ket{\psi,\phi} \lra \psi^1\phi^2 E I\sk^1 = \psi^1\phi^2 E I\sk^2 = 
\psi^1\phi^2 J. 
\end{equation} 
Consistency in this formulation is ensured by the results
\begin{equation}
J^2=-E \quad \mbox{and} \quad
J=JE=EJ. 
\end{equation}

The action of the individual Pauli matrices now becomes, for example, 
\begin{equation}
\hat{\sigma}_k \otimes \hat{I} \, \ket{\psi} \lra - I\sigma_k^1 \psi J
\end{equation}
where $\hat{I}$ is the $2\times 2$ identity matrix.  A similar result
holds for the second particle space.  The action on the right-hand
side keeps us in the space of correlated products of even elements of
$\clg_3$.  The quantum inner product is replaced by the operation
\begin{equation} 
\braket{\psi}{\phi} \lra (\psi,\phi)_s = 2\la \phi E\tilde{\psi}
\ra - 2\la \phi J \tilde{\psi} \ra i.
\label{2ptinner}
\end{equation} 
The factor of $E$ in the real part is not strictly necessary as it is
always present in the spinors, but including it does provide a neat
symmetry between the real and imaginary parts.  The factor of 2 is
included to ensure complete consistency with the standard quantum
inner product.  (In the general $n$-particle case a factor of
$2^{n-1}$ is required.)

In Section \ref{sd} we found that a general 2-particle
wavefunction can be written in the form of Eq.~(\ref{Schmidt}).  To
find the geometric algebra form of this we first define the spinor
\begin{equation}
\psi(\theta, \phi) = e^{-\phi I\sigma_3/2}e^{-\theta I\sigma_2/2} .
\end{equation}
We also need a representation of the orthogonal state to this, which
is
\begin{align} 
\spinor{\sin(\theta/2) e^{-i\phi/2}}{-\cos(\theta/2) e^{i\phi/2}}
\lra \, &
\sin(\theta/2) e^{-I\sigma_3 \phi/2} + \cos(\theta/2) I\sj
e^{I\sigma_3 \phi/2} \nonumber \\
= & \psi(\theta, \phi) I \sj.
\end{align}
It is a straightforward exercise to confirm that this state is
orthogonal to $\psi(\theta, \phi)$, as required.  We can now construct
the MSTA version of the Schmidt decomposition.  We replace
Eq.~(\ref{Schmidt}) with
\begin{align}
\psi =& \rho^{1/2} \Bigl( \cos(\alpha/2) \psi^1(\theta_1,\phi_1)
\psi^2(\theta_2,\phi_2) e^{J \tau/2} \nonumber \\ 
& +  \sin(\alpha/2) \psi^1(\theta_1,\phi_1) \psi^2(\theta_2,\phi_2)
I\sj^1 I\sj^2 e^{-J \tau/2} \Bigr) e^{J \chi}E  \nonumber \\
=& \rho^{1/2} \psi^1(\theta_1,\phi_1) \psi^2(\theta_2,\phi_2) e^{J
\tau/2} \left( \cos(\alpha/2) +  \sin(\alpha/2) I\sj^1 I\sj^2  \right)
e^{J \chi} E. 
\end{align}
If we now define the individual rotors
\begin{equation}
R = \psi(\theta_1,\phi_1) e^{I\sk \tau/4}, \quad
S = \psi(\theta_2,\phi_2) e^{I\sk \tau/4}, 
\end{equation}
then the wavefunction $\psi$ can be written compactly as
\begin{equation}
\psi =  \rho^{1/2} R^1 S^2  \left( \cos(\alpha/2) +  \sin(\alpha/2)
I\sj^1 I\sj^2  \right) e^{J \chi} E. 
\label{finpsi}
\end{equation}
This gives a neat, general form for an arbitrary 2-particle state.  In
particular, all reference to the tensor product has been dropped in
favour of the somewhat simpler geometric product.  The degrees of
freedom are held in an overall magnitude and phase, two separate
rotors in the individual particle spaces, and a single entanglement
angle $\theta$.  In total this gives 9 degrees of freedom, so one of
them must be redundant.  This redundancy is in the single-particle
rotors.  If we take
\begin{equation}
R \mapsto R e^{I\sk \beta}, \quad S \mapsto S e^{-I\sk \beta}
\end{equation}
then the overall wavefunction $\psi$ is unchanged.  In practice this
redundancy is not a problem, and the form of~(\ref{finpsi}) turns out
to be extremely useful.

The GA form of the Schmidt decomposition in~(\ref{finpsi}) is very
suggestive of a more general pattern.  To the left we have rotation
operators in each of the individual spaces.  In one sense the rotors
$R^1 S^2$ can be viewed as representing the nearest direct product
(separable) state.  Next comes a term describing the 2-particle
entanglement.  The generalisation seems fairly clear.  For a
3-particle system we expect to see terms describing the various
2-particle entanglements, followed by a term for the 3-particle
entanglement.  Finding precisely the optimal decomposition along these
lines is an open problem, but the GA formalism has suggested an
approach to the general problem of classifying multiparticle
entanglement which has not been tried before.

\subsection{2-Particle Observables}

We can start to appreciate the utility of the form of~(\ref{finpsi})
by studying the 2-particle observables.  These go as, for example
\begin{equation}
\bra{\psi} \, \hat{\sigma}_k \otimes \hat{I} \, \ket{\psi} 
\lra (\psi, -I\sigma_k^1 \psi J)_s = - 2 I\sigma_k^1 \dt (\psi J
\tilde{\psi}) 
\end{equation}
and
\begin{equation}
\bra{\psi} \, \hat{\sigma}_j \otimes \hat{\sigma}_k \, \ket{\psi}  
\lra (\psi, - I\sigma_j^1 \, I\sigma_k^2 \psi)_s = - 2 (I\sigma_j^1 \,
I\sigma_k^2) \dt (\psi E \tilde{\psi}).
\end{equation}
All of the observables one can construct are therefore contained in
the multivectors $\psi E \tilde{\psi}$ and $\psi J \tilde{\psi}$.
This is true in the general $n$-particle case, and is a major strength
of the MSTA approach.

To study the form of the observables we first simplify slightly and set
$\rho=1$.  We find that (using $E \tilde{E}=EE=E$)
\begin{align}
\psi E \tilde{\psi} 
&= R^1 S^2 \bigl(\cos(\alpha/2) + \sin(\alpha/2) I\sigma_2^1
I\sigma_2^2 \bigr) E \nonumber \\
& \qquad \bigl(\cos(\alpha/2)+ \sin(\alpha/2) I\sigma_2^1 I\sigma_2^2 \bigr)
\tilde{R}^1 \tilde{S}^2 \nonumber \\
&= R^1 S^2 \bigl(1+ \sin(\alpha)I\sigma_2^1 I\sigma_2^2 \bigr)E
\tilde{R}^1 \tilde{S}^2
\end{align}
Substituting in the form of $E$ from Eq.~(\ref{E}) gives
\begin{equation} 
\psi E \tilde{\psi} = \half R^1S^2 \Bigl(1-I\sigma_3^1 I\sigma_3^2 +
\sin(\alpha)(I\sigma_2^1 I\sigma_2^2-I\sigma_1^1 I\sigma_1^2) \Bigr)
\tilde{R}^1 \tilde{S}^2.
\end{equation}
To make this result clearer we introduce the notation
\begin{equation} 
A_k = RI\sigma_k\tilde{R}, \quad
B_k = SI\sigma_k\tilde{S}
\end{equation} 
so that
\begin{equation} 
\psi E \tilde{\psi} =
\half (1-A_3^1B_3^2)  + \half \sin (\alpha) (A_2^1B_2^2 - A_1^1B_1^2).
\end{equation} 
From this we see that
\begin{equation} 
\la\psi E \tilde{\psi}\ra=\half.
\end{equation}
This factor of one-half is absorbed by the factor of 2 in the
definition of the quantum inner product~(\ref{2ptinner}) and shows
that the state is correctly normalised to 1.  The 4-vector part of the
observable is more interesting, as it contains combinations of $A_1,
A_2, B_1, B_2$, none of which are accessible to measurement in the
single-particle case (as they are not phase invariant).  This is one
place where differences between classical and quantum models of spin
start to emerge.

The second observable to form from the 2-particle state $\psi$ is
$\psi J\tilde{\psi}$, which is given by  
\begin{align}
\psi J\tilde{\psi}
&= R^1 S^2 \bigl( \cos(\alpha/2) + \sin(\alpha/2)I\sigma_2^1
I\sigma_2^2 \bigr) J 
\nonumber \\
& \qquad \bigl( \cos(\alpha/2) + \sin(\alpha/2)I\sigma_2^1
I\sigma_2^2 \bigr)  \tilde{R}^1 \tilde{S}^2  \nonumber \\
&=\half R^1 S^2 \bigl( \cos^2(\alpha/2)-
\sin^2(\alpha/2)\bigr)(I\sigma_3^1+I\sigma_3^2)  \tilde{R}^1
\tilde{S}^2  \nonumber \\ 
&=\half \cos(\alpha) (A_3^1+B_3^2 ) 
\label{pJp}
\end{align}
This result extends the definition of the polarisation bivector to
multiparticle systems.  An immediate consequence of this definition is
that the lengths of the bivectors are no longer fixed, but instead
depend on the entanglement.

\subsection{The Density Matrix}

The density matrix for a normalised 2-particle pure state can be
expanded in terms of products of Pauli matrices as 
\begin{equation}
\hat{\rho} = \ket{\psi} \bra{\psi} =
\frac{1}{4} \bigl( \hat{I} \otimes \hat{I} + a_k \, \hat{\sigma}_k
\otimes \hat{I} + b_k \, \hat{I} \otimes \hat{\sigma}_k 
+ c_{jk} \, \hat{\sigma}_j \otimes \hat{\sigma}_k   \bigr).
\end{equation}
The various coefficients are found by forming, for example
\begin{equation}
a_k = \bra{\psi} \, \hat{\sigma}_k \otimes \hat{I} \, \ket{\psi} 
= -2 I\sigma_k^1 \dt (\psi J \tilde{\psi}) 
\end{equation}
It follows that all of the degrees of freedom present in the density
matrix are contained in the multivector observables $\psi E
\tilde{\psi}$ and $\psi J \tilde{\psi}$.  For mixed states we simply
add the weighted values of these observables formed from the pure
states.  This picture is quite general and works for any number of
particles.  One small complication is that the terms in $\psi J
\tilde{\psi}$ are anti-Hermitian, whereas the density matrix is
Hermitian.  One way round this is to correlate all of the
pseudoscalars together and map all bivectors back to their dual
vectors~\cite{hav_dor1} .  One can often ignore this feature, however,
and work directly with the observables $\psi E \tilde{\psi}$ and
$\psi J \tilde{\psi}$.

An advantage of this way of encoding the density matrix is that the
partial trace operation to form the reduced density matrix simply
consists of throwing away any terms in the observables coming from
spaces where the state is unknown.  For example, taking the 2-particle
entangled state~(\ref{finpsi}) and tracing out the degrees of freedom
in space 2 just leaves
\begin{equation}
\hat{\rho} = \half (1 + P_k \hat{\sigma}_k), \quad P_k = (-I\sigma_k) \dt
\bigl(\cos(\alpha) R I \sigma_3 \tilde{R} \bigr).
\end{equation}
This shows that the effect of the entanglement is to reduce the
expectation value for the polarisation from 1 to $\cos(\alpha)$, but
leave the direction of polarisation unchanged.  For 2-particle pure
states we also see that the polarisation vector is the same length for
both particles, so each particle is effected equally by any
entanglement which is present.  For higher particle number or mixed
states the effects of entanglement are more complicated, though the
formula 
\begin{equation}
P_k = - 2^{n-1} (I\sigma_k^a) \dt (\psi J \tilde{\psi})
\end{equation}
holds whenever we form the reduced density matrix for particle $a$
from a larger, entangled state.

Our simple 2-particle system exhibits one of the basic results of
quantum theory.  When a system entangles with a second, unknown system
(usually the environment), the state of the system of interest can no
longer be known for certain and we are forced to adopt a density
matrix viewpoint.  That is, entanglement with the environment leads to
decoherence and loss of information.

A useful application of the preceding is to the overlap probability
for the inner product of two states.  Given two normalised states we
have
\begin{equation}
P(\psi, \phi) = |\la \psi | \phi \ra|^2 = \mbox{tr} (\hat{\rho}_\psi
\hat{\rho}_\phi).  
\end{equation}
The degrees of freedom in the density matrices are contained in $\psi
E \tilde{\psi}$ and $\psi J \tilde{\psi}$, with equivalent expressions
for $\phi$.  One can then show that the probability is given by the
compact expression
\begin{equation}
P(\psi, \phi) = \la (\psi E \tilde{\psi}) (\phi E \tilde{\phi}) \ra -
\la (\psi J \tilde{\psi}) (\phi J \tilde{\phi}) \ra .
\label{2ptin}
\end{equation}
This formula holds in the $n$-particle case as well, except for the
presence of an additional factor of $2^{n-2}$ to give the correct
normalisation.  This compact expression is a unique feature of the
MSTA approach.

As a check on the preceding, suppose we have two separable states
\begin{equation}
\psi = R^1 S^2 E, \quad \phi = U^1 V^2 E
\end{equation}
with
\begin{equation}
\psi E \tilde{\psi} = \half(1-A^1 B^2), \quad 
\phi E \tilde{\phi} = \half(1-C^1 D^2) .
\end{equation}
We find that
\begin{align} 
P(\psi, \phi) 
&= \qrt \la (1-A^1 B^2) (1-C^1 D^2) - (A^1 + B^2) (C^1 + D^2)
\ra \nonumber \\
&= \qrt ( 1 + A \dt C \, B \dt D - A \dt C - B \dt D) \nonumber \\
&= \half(1- A \dt C ) \, \half (1- B \dt D)
\end{align}
which shows that the probability is the product of the separate
single-particle probabilities.  If one of the states is entangled this
result no longer holds.

\subsection{Example -- The Singlet State}

As a simple example of the some of the preceding ideas, consider the
spin-singlet state
\begin{equation}
\ket{\psi} = \frac{1}{\sqrt{2}} (\ket{01} - \ket{10})
\lra \psi = \frac{1}{\sqrt{2}} (I\sj^1 - I \sj^2) E.
\end{equation}
This state is maximally entangled ($\alpha=\pi/2$), and isotropic.
Forming the two observables we find that
\begin{equation}
\psi E \tilde{\psi} = \half (1 + I \sigma_k^1 \, I \sigma_k^2 )
\end{equation}
and
\begin{equation}
\psi J \tilde{\psi} = 0.
\end{equation}
It follows that the reduced density matrix for either particle space
is simply one-half of the identity matrix, and so all directions
are equally likely.  If we align our measuring apparatus along some
given axis and measure the state of particle one then both up and down
have equal probabilities of one-half.

Suppose now that we construct a joint measurement on the singlet
state.  We can model this as the overlap probability between $\psi$
and the separable state
\begin{equation}
\phi = R^1 S^2 E.
\end{equation}
Denoting the spin directions by
\begin{equation}
R I \sk \tilde{R} = P, \quad S I \sk \tilde{S}=Q,
\end{equation}
we find that, from~(\ref{2ptin})
\begin{align}
P(\psi,\phi) 
&= \la \half(1-P^1 Q^2) \half(1 + I \sigma_k^1 \, I \sigma_k^2 ) \ra
\nonumber \\ 
&= \qrt (1 - P \dt (I\sigma_k) \, Q \dt (I\sigma_k)) \nonumber \\
&= \qrt(1 - \cos\theta) 
\end{align}
where $\theta$ is the angle between the spin bivectors $P$ and $Q$.
So, for example, the probability that both measurements result in the
particles having the same spin ($\theta = 0$) is zero, as expected.
Similarly, if the measuring devices are aligned, the probability that
particle one is up and particle two is down is one-half, whereas if
there was no entanglement we should get the product of the separate
single particle measurements (resulting in $1/4$).

It is instructive to see how all of strange quantum entanglement
results for the singlet state are contained in the 4-vector part of
the observables.  This reveals some of the complex geometry associated
with multiparticle quantum mechanics.  And this is only for 2-particle
systems!  Most proposals for quantum computers have in mind a far
greater number of entangled qubits.  We hope that this paper has
demonstrated some of the potential power of geometric algebra for
helping to navigate through these large Hilbert spaces.

\section{Acknowledgements}

Rachel Parker is supported by the Cambridge Commonwealth Trust and the
Kerry Packer Scholarship Foundation.  Chris Doran is supported by the
EPSRC.  The authors thank Anthony Lasenby and Timothy Havel for
helpful discussions.


\begin{thebibliography}{10}

\bibitem{lomo01}
S.J. Lomonaco.
\newblock An entangled tale of quantum entanglement.
\newblock quant-ph/0101120.

\bibitem{DGL93-states}
C.J.L. Doran, A.N. Lasenby, and S.F. Gull.
\newblock States and operators in the spacetime algebra.
\newblock {\em Found. Phys.}, 23(9):1239, 1993.

\bibitem{DGL95-elphys}
C.J.L Doran, A.N. Lasenby, S.F. Gull, S.S. Somaroo, and A.D. Challinor.
\newblock Spacetime algebra and electron physics.
\newblock {\em Adv. Imag. \& Elect. Phys.}, 95:271, 1996.

\bibitem{SLD99}
S.S. Somaroo, A.N. Lasenby, and C.J.L. Doran.
\newblock Geometric algebra and the causal approach to multiparticle quantum
  mechanics.
\newblock {\em J. Math. Phys.}, 40(7):3327--3340, 1999.

\bibitem{bell-spk}
J.S. Bell.
\newblock {\em Speakable and unspeakable in quantum mechanics}.
\newblock Cambridge University Press, 1987.

\bibitem{ekert95}
A.~Ekert and P.L. Knight.
\newblock Entangled quantum systems and the {S}chmidt decomposition.
\newblock {\em Am. J. Phys.}, 63(5):415, 1995.

\bibitem{paz00}
J.P. Paz and W.H. Zurek.
\newblock Environment-induced decoherence and the transition from quantum to
  classical.
\newblock quant-ph/0010011.

\bibitem{hes67}
D.~Hestenes.
\newblock Real spinor fields.
\newblock {\em J. Math. Phys.}, 8(4):798, 1967.

\bibitem{hes71}
D.~Hestenes.
\newblock Vectors, spinors, and complex numbers in classical and quantum
  physics.
\newblock {\em Am. J. Phys.}, 39:1013, 1971.

\bibitem{hes751}
D.~Hestenes and R.~Gurtler.
\newblock Consistency in the formulation of the {D}irac, {P}auli and
  {S}chr{\"{o}}dinger theories.
\newblock {\em J. Math. Phys.}, 16(3):573, 1975.

\bibitem{hes-nf1}
D.~Hestenes.
\newblock {\em New Foundations for Classical Mechanics (Second Edition)}.
\newblock Kluwer Academic Publishers, Dordrecht, 1999.

\bibitem{DL-course}
C.J.L. Doran and A.N. Lasenby.
\newblock Physical applications of geometric algebra.
\newblock Cambridge University Lecture Course. Lecture notes available from
  \texttt{http://www.mrao.cam.ac.uk/$\sim$clifford/ptIIIcourse}.

\bibitem{hav_dor1}
T.F. Havel and C.J.L. Doran.
\newblock Geometric algebra in quantum information processing.
\newblock quant-ph/0004031, \textit{AMS Contemporary Math series}, to appear.

\end{thebibliography}
\end{document}